\documentstyle[12pt]{article}

\begin{document}
\def\bib#1{[{\ref{#1}}]}
\def\at{\tilde{a}}

\begin{titlepage}
         \title{ Exact Gravitational Gauge Structures  \\
                 and the Dirac Equation \footnote{Research supported by 
                 Ministero dell'Universit\'a e della Ricerca Scientifica}}

\author{G. Lambiase$^{a,b}$\thanks{Fax: +39-89-965275. 
E-mail:lambiase@vaxsa.csied.unisa.it}
and~~~G.Papini$^{c,}$\thanks{Tel: (306) 585-4204.
Fax: (306) 585-4894. E-mail: papini@cas.uregina.ca}  \\
{\em  $^a$ Dipartimento di Fisica Teorica e S.M.S.A.}\\ 
{\em  Universit\`a di Salerno, 84081 Baronissi (Salerno), Italy.}\\
{\em  $^b$ Istituto Nazionale di Fisica Nucleare, Sezione di Napoli,}\\
{\em Mostra d'Oltremare pad. 19, 80125 Napoli, Italy.}\\
{\em $^c$ Department of Physics, University of Regina}, \\
{\em Regina, Sask. S4S 0A2, Canada.}}
              \date{\empty}
              \maketitle

              \begin{abstract}
Exact gauge structures arise in the evolution of spin-$\frac{1}{2}$
particles in conformally flat space-times. The corresponding
Berry potentials can be Abelian or non-Abelian depending on the
mass degeneracy of the system considered. Examples include de Sitter 
universes and maximal acceleration.
	      \end{abstract}

\thispagestyle{empty}
\vspace{20. mm}
\vspace{5. mm}
              \vfill
	      \end{titlepage}
Quantum phases of the type discussed by Berry \cite{BER} arise from
the cyclic variation of a system's parameters in parameter space.
Thus the gauge potentials that represent a geometrical structure in
parameter space, induce a corresponding structure in Hilbert space.
Since gravitational fields endow space-time with geometrical structure,
the latter should also affect Hilbert space when parameter space and
space-time coincide. This was argued in a recent past \cite{CAI}
and geometrical phases of this type were indeed found \cite{YQC} and
used to discuss a variety of gravitational and inertial phenomena
from a unified point of view \cite{PAP}.
Though derived using a different procedure than the original one by Berry,
these phases are deeply geometrical and adapt themselves more
readily to relativistic problems.

In the particular context of the Dirac equation, two geometrical phases 
were discovered \cite{GPA}. One of these, exact, records exclusively
the history of the particle spin in space-time. The second, exact only
to first order in the metric deviation, refers to the space-time
development of the particle independently of its spin \cite{YQC}.
Both were more directly identifiable from the point of view of the
non-inertial observer, but could still be expressed in covariant form.

The purpose of this work is to present one more exact geometrical phase, 
still in the context of spin-$\frac{1}{2}$ particles, but from 
the viewpoint
of an inertial observer. It refers to conformally flat space-times,
which describe universes of constant curvature and play important roles
from cosmology to hadronic confinement. The connection between
confinement and conformally flat solutions of Einstein equations 
resides in the properties of the Einstein tensor with cosmological
term. Under conformally flat transformations the tensor takes the form
of the improved energy-momentum tensor $\Theta_{\alpha\beta}$ of
Callan, Coleman and Jackiw \cite{CCJ} whose behaviour determines the
local distribution of vacuum energy.
It follows, for instance, that singular solutions of the equation 
$\Theta_{\alpha\beta}=0$ which yield Abelian instantons, are also 
solutions of Einstein equations. Similar situations exist for (Abelian)
merons and elliptic solutions that interpolate between merons and
instantons \cite{AGP}. The corresponding gravitational fields represent 
macro-universes, or micro-universes, according to appropriate choices
of the constants involved. Micro-universes yield hadron bags
\cite{SAL}.

A more unusual conformal field, also relevant to confinement, will be
discussed later. It does not correspond to any particular solution of
Einstein equations, but to the interesting suggestion that proper
accelerations have un upper limit.

In what follows, calculations refer to a gravitational field of metric
\begin{equation}
g_{\mu\nu}=\sigma^2(x)\eta_{\mu\nu}\,{,}
\end{equation}
where $\eta_{\mu\nu}$ is a Minkowski metric of signature $-2$, and
to the covariant Dirac equation
\begin{equation}
(i\hbar\gamma^{\mu}(x){\cal D}_{\mu}-m)\psi(x)=0\,{,}
\end{equation}
where ${\cal D}\equiv \nabla_{\mu}+\Gamma_{\mu}(x)$. The operator of
covariant differentiation is $\nabla_{\mu}$, $\Gamma_{\mu}$ the 
spinorial connection and $\gamma^{\mu}(x)$ are position-dependent
matrices that satisfy the anti-commutation relations 
$\{\gamma^{\mu}(x),\gamma^{\nu}(x)\}=2g_{\mu\nu}(x)$. The geometrical
phase can be derived by referring (2) to a set of inertial observers
by means of a tetrad field $e_{\mu}^{\,\,a}(x)$, where Latin indices
are tetrad indices \cite{GIA}. Then the total connection is represented 
by 
\begin{equation}
\tilde{\Gamma}_{\mu}=\frac{1}{2}\sigma^{ab}\omega_{\mu ab}\,{,}
\end{equation}
where
\begin{equation}
\sigma^{ab}=\frac{1}{4}\,[\gamma^a,\gamma^b]
\end{equation}
\begin{equation}
\omega_{\mu\,\,\,\,b}^{\,\,\,\,a}=
(\Gamma^{\lambda}_{\mu\nu}e_{\lambda}^{\,\,\,\,a}-
\partial_{\mu}e_{\nu}^{\,\,\,\,a})e^{\nu}_{\,\,\,\,b}
\end{equation}
\begin{equation}
\Gamma^{\lambda}_{\mu\nu}=\frac{1}{2}
g^{\lambda\alpha}(g_{\alpha\mu,\nu}+g_{\alpha\nu,\mu}-g_{\mu\nu,\alpha})
\end{equation}
\begin{equation}
\gamma^{\mu}(x)=e^{\mu}_{\,\,\,\,a}\gamma^a,\quad 
e^{\mu}_{\,\,\,\,a}e_{\mu}^{\,\,\,\,b}
=\delta_{a}^{\,\,\,\,b}\,{.}
\end{equation}
The form of (1) determines the tetrad field
\begin{equation}
e_{\mu}^{\,\,\,\,a}(x)=\sigma(x)\delta_{\mu}^{\,\,\,\,a}\,{.}
\end{equation}
Also from (3)-(8) one finds
\begin{equation}
\tilde{\Gamma}_{\mu}=\sigma^{ab}\eta_{a\mu}(\ln\sigma)_{,b}\,{.}
\end{equation}
For the set of inertial observers along the particle worldline, 
Eq. (2) becomes
\begin{equation}
\left[i\hbar\gamma^a\partial_a+i\frac{3\hbar}{2}\gamma^a(\ln\sigma)_{,a} -m
\right]\psi(x)=0\,{,}
\end{equation}
with solution
\begin{equation}
\psi(x)=\exp\left[-\frac{3}{2}\int_{x_0}^{x}(\ln\sigma)_{,a}dx^a\right]
\psi_0(x)\,{,} 
\end{equation}
where $\psi_0(x)$ satisfies the free Dirac equation in Minkowski space.
Henceforth all quantities refer to inertial observers and Latin indices 
are replaced by Greek ones for convenience. The integral
\begin{equation}
\phi=-\frac{3}{2}\int_{x_0}^{x}(\ln\sigma)_{,\mu}dx^{\mu}=
-\frac{3}{2}\int_{s_0}^{s}(\ln\sigma)_{,s}ds\,{,}
\end{equation}
carries all the information regarding the gravitational field and 
represents the desired result. It bears a close resemblance to the Dirac 
phase factor for electromagnetic fields and the vector
\begin{equation}
B_{\mu}=i\frac{3\hbar}{2}(\ln\sigma)_{,\mu}
\end{equation}
is the corresponding potential. Eq. (12) does not in general represent
a geometrical phase because extension to a closed path yields a vanishing 
result whenever $\sigma(x)$ is well behaved. It becomes a true 
geometrical phase if $\sigma(x)$ vanishes or has singularities.
This new phase does not, of course, preclude the existence of other
geometrical phases or the use of those already discussed in the
literature. Its existence merely underlines how in relativity
different approaches may bring out different aspects of a problem.

The result can be extended by using the method of Ref. \cite{CAI}
which is based on the relativistic equation
\begin{equation}
i\frac{\partial\Phi}{\partial\tau}=H[\lambda_a(\tau)]\Phi\,{,}
\quad a=1,2,\ldots,k
\end{equation}
rather than the usual Schr\"{o}dinger equation. 
In Eq. (14) $\tau$
is a relativistic invariant parameter that describes the evolution generated 
by the relativistic operator $H=-(\Box-U(x))$. By restricting the solutions
of (14) to those of the type $\Phi=e^{im^2\tau}\phi(x)$, one finds
\begin{equation}
H\phi(\lambda)=-m^2\phi(\lambda)\,{,}
\end{equation}
where $m$ is the mass of a particle. The Dirac equation and most other
wave equations can be re-cast in the form of Eq. (15). Cyclic evolution 
of the parameters $\lambda_a(\tau)$ leads to a covariant generalization 
of Berry's phase
\begin{equation}
\gamma(C)=i\oint<\phi|\frac{\partial}{\partial\lambda_a}|\phi>d\lambda_a\,{,}
\end{equation}
provided the initial pure mass square eigenstate does not change
during the evolution. A major point of this generalization is that
it is now possible to identify parameter space with ordinary
space-time by setting $k=4$ and $\lambda_a\equiv x_{\mu}(\tau)$. In 
particular, Eq. (10) can be re-cast in the form (15) by introducing a 
new spinor $\psi^{\prime}$ defined 
by $\psi\equiv (i\hbar\gamma^{\mu}\partial_{\mu}
+\gamma^{\mu}B_{\mu}+m)\psi^{\prime}$ and (16) then yields (12) as expected.
In addition the gauge structures can be generalized to the non-Abelian case 
when the eigenvalues of Eq. (15) are $N$-fold degenerate
\begin{equation}
H(\tau)\psi_i^a(\tau)=-m^2_a\psi^a_i(\tau) \quad i=1,2,\ldots,N\,{.}
\end{equation}
The $SU(N)$ symmetry of these states then leads to Yang-Mills
fields, while the corresponding potentials are given by \cite{CAI}
\begin{equation}
A^a_{\mu ij}=<\psi_j^a|\frac{\partial\psi_i^a}{\partial x^{\mu}}>\,{.}
\end{equation}

We now return to Eq. (12). When extended to a closed path, the integral
may be evaluated by a variety of means. If it can be extended to the 
complex domain, then one may use the formula
\begin{equation}
\oint(\ln\sigma)_{,z}dz=2\pi i(N-P)\,{,}
\end{equation}
where $N$ and $P$ represent the numbers of zeros and poles od $\sigma(z)$
within the closed path. This information, in fact, characterizes conformally
flat fields. In particular, de Sitter space-time solutions are of the form
\begin{equation}
\sigma(x)=\left(\Lambda\frac{x^2-a^2}{a}\right)^{-1}\,{,}
\end{equation}
where $\Lambda$ is a constant, $a_{\mu}$ a constant vector and 
$a^2=a_{\mu}a^{\mu}$. 
It is solution (20) that substituted in (10), provides a 
time-dependent potential term in which the particle is embedded. Solution
(20) is in fact an Abelian instanton in the context of confinement.
In this instance the surface $x^2-a^2=0$ is singular and $\sigma(x)$
has poles at the point of intersection of the particle worldline
with this surface. The corresponding change in phase along paths through a 
point of intersection follows from the formula $(\ln\sigma(s))_{,s}=
(\ln|\sigma(s)|)_{,s}+i\pi\delta(\sigma(s))\sigma^{'}(s)$ and is
$i\pi\gamma$, where $\gamma$ is the sign function. Though de Sitter
spaces are singularity free, it is an interesting consequence
of (12) that their singularities in Minkowski space are physically
meaningful. On the other hand, these singularities are removable
only in non-inertial frames, in which instance the Dirac equation
reverts to its general form (2) with no geometrical phase of the type (12).

Merons, with no singularities in Minkowski space, have no phases of the type
under discussion for any closed space-time paths in this space.

A particular interesting, though unusual example of a conformal field is
the one introduced by Caianiello and collaborators in connection with the 
notion of maximal acceleration \cite{LLO}. Based on the standard 
interpretation of quantum mechanics and eight-dimensional phase space, this 
theory implies that a particle of mass $m$ accelerating along its worldline,
in the absence of gravity, behaves as if embedded in an effective gravitational
field 
\begin{equation}
g_{\mu\nu}=(1-\frac{c^4|\ddot{x}|^2}{A^2})\eta_{\mu\nu}\,{,}
\end{equation}
where $A=2mc^3/\hbar$ is the maximal acceleration of the particle
and $|\ddot{x}|^2\equiv |\eta_{\mu\nu}\ddot{x}^{\mu}\ddot{x}^{\nu}|$
is the square length of its acceleration four-vector. Similar results were
also found by Papini and Wood \cite{PWD} starting from an entirely 
geometrical viewpoint.

Among the various consequences of Caianiello's theory is confinement, the 
bag being produced by the acceleration of the quarks themselves \cite{SCA}.
The present analysis which refers Eq. (10) in the presence of the field (21) 
to a set of inertial observers at all points along the classical worldline
of the particle, uncovers other features of the model. First, acceleration
of systems with degeneracy would also generate non-Abelian potentials
given by (18) above. Second, for values of $|\ddot{x}|<A/c^2$
($\sigma\ne 0$),
which result in no poles or zeros being enclosed by $C$,  
an effect could still be observed in principle by interferometric means using
the phases previously discussed \cite{CAI}-\cite{GPA},\cite{GFS}. 
Third for values of 
$\ddot{x}$ such that $c^2|\ddot{x}|=A$, $\sigma$ vanishes on the particle
worldline and the contribution of any of these zeros to the phase is
$3\pi/2$ in absolute value. 
It is doubtful whether this change in phase could be observed by 
interferometric means, given the drastic nature of the transition through
$\sigma=0$, as indicated below. Alternatively a  path may be chosen
to link the worldline of the particle along which $\sigma=0$. This
phenomenon is analogous to the occurrence of a magnetic pole in 
Dirac's theory.
The flux of $B_{\mu}$ through any open surface bound by the path is 
then quantized
in units of $3\pi$. Finally, the case of total phase quantization
in the presence of external electromagnetic fields is also interesting.
In this instance the total phase must be $2\pi n$. Since the minimum
contribution to it from $B_{\mu}$ is $3\pi$, the magnetic flux 
contained in the world tube must jump by $3\pi c\hbar/e$ at each 
transition point $\sigma=0$, when $n$ is constant.

Though accelerations of the order of the maximal acceleration may already 
be present in electron-positron collisions with the formation of a $Z_0$
nearly at rest, the condition $\sigma=0$, if attainable, would have additional
interesting consequences. For one, the sign of the effective metric
would change and the light cone at the particle would rotate, effectively
disconnecting the particle from the outside world.

In conclusion, exact Abelian and non-Abelian gauge structures are generated 
in the dynamics of spin-$\frac{1}{2}$ particles in conformally flat space-times
and non-Abelian potentials arise when the particle's mass
spectrum is $N$-fold degenerate. Since space-time, rather than parameter space,
is the space where the system's development takes place, the Abelian fields
are true space-time dependent fields and the non-Abelian ones are true
Yang-Mills fields with $SU(N)$ symmetry. Characteristics of the gauge
fields are related to flat space-time zeros or singularities of the
conformal factor $\sigma(x)$. These are the points around which phase changes
take place and it is around these points that the system organizes the 
information geometrically.

\bigskip
\begin{centerline}
{\bf Acknowledgments}
\end{centerline}

G.P. gladly acknowledges the continued research support 
of Dr. K. Denford, Dean of Science, University of Regina.

G.L. wishes to thank Dean Denford for his kind hospitality during a stay
at the University of Regina.
 
\newpage
\begin{centerline}
{\bf REFERENCES}
\end{centerline}

\vfill

\end{document}